\documentclass[12pt,thmsa]{article}

\input{tcilatex}
\begin{document}

\author{Harald Grosse$^a$ \and Karl-Georg Schlesinger$^b$ \qquad \\
$^a$Institute for Theoretical Physics\\
University of Vienna\\
Boltzmanngasse 5\\
A-1090 Vienna, Austria\\
e-mail: grosse@doppler.thp.univie.ac.at\\
$^b$Erwin Schr\"{o}dinger Institute for Mathematical Physics\\
Boltzmanngasse 9\\
A-1090 Vienna, Austria\\
e-mail: kgschles@esi.ac.at}
\title{Spinfoam models for $M$-theory }
\date{}
\maketitle

\begin{abstract}
We use the approach to generate spin foam models by an auxiliary field
theory defined on a group manifold (as recently developed in quantum gravity
and quantization of $BF$-theories) in the context of topological quantum
field theories with a 3-form field strength. Topological field theories of
this kind in seven dimensions are related to the superconformal field
theories which live on the worldvolumes of fivebranes in $M$-theory. The
approach through an auxiliary field theory for spinfoams gives a topology
independent formulation of such theories.
\end{abstract}

\section{Introduction}

In the field of quantum gravity and quantization of $BF$-theories there has
recently been given a formulation of spinfoam models for such theories in
terms of an auxiliary field theory on a group manifold which gives these
theories a completely topology independent form and, hence, allows a sum
over different topologies in the path integral (see \cite{LPR}, \cite{RR}
and the literature cited therein). The simplest example (discussed in \cite
{LPR})\ is $BF$-theory in two dimensions with an $SU\left( 2\right) $
connection, i.e. we have a scalar field $B$, an $SU\left( 2\right) $ valued
1-form $A$ with $F$ its 2-form field strength and the action 
\[
S\left( A,B\right) =\int Tr\left( BF\right) 
\]
In this paper, we utilize the approach through an auxiliary field theory on
a group manifold to give a topology independent formulation to what we will
call $BH$-theories, in the sequel. By this, we mean a topological field
theory with a field $B$ which is in $d$ dimensions given by a $\left(
d-3\right) $-form (and otherwise behaves completely analogous to the $B$%
-field in $BF$-theory, hence the terminology) and an $SU\left( 2\right) $
valued 2-form $G$ with a 3-form field strength $H$ with 
\[
H=dG
\]
and the action 
\[
S\left( G,B\right) =\int Tr\left( BH\right) 
\]
Obviously, the simplest case is the three dimensional one which we will
study in detail in this paper. Besides this, we will discuss how the
generalization to higher dimensions, especially to seven dimensions, looks
like. Topological field theories with a 3-form field strength in seven
dimensions are of special interest because they are related to the six
dimensional superconformal field theories living on the worldvolumes of
fivebranes in $M$-theory (see \cite{Dij}). This correspondence parallels the
one between two dimensional superconformal and three dimensional topological
field theories which is of considerable importance in perturbative string
theory.

The approach through an auxiliary field theory on a group manifold allows to
give a formulation of such topological field theories with a 3-form field
strength which is completely independent of topology and involves a natural
prescription for a sum over topologies in the path intergral. The main
ingredient in generalizing the approach through an auxiliary field theory
from $BF$ to $BH$ theories will be the observation that in the
discretization of a theory with a 3-form as compared to a 2-form field
strength connections no longer live on edges but on two dimesional faces.
This is similar to the extended lattice gauge theories introduced in \cite
{GS1} (with the difference that here the colouring is taken directly from
the group $SU\left( 2\right) $ instead of the category of representations)
and is related to the fact that 3-form theories relate in the continuum case
to gerbes on the background manifold instead of vector bundles (see \cite{BM}%
).

\bigskip

\section{The three dimensional case}

We start by briefly discussing the topological content of $BH$ theory (in
complete analogy to the treatment of $BF$ theory in \cite{LPR}). This part
is, of course, dependent on the topology of the background manifold. In the
second part of this section, we will then consider the approach by an
auxiliary field theory on a group manifold. By the definition of the 3-form
curvature of connections on gerbes (see \cite{BM}) it follows that a
discrete approximation of the action 
\[
S\left( G,B\right) =\int Tr\left( BH\right) 
\]
is given by 
\[
S\left( B_s,U_f\right) =\sum_sTr\left( B_sU_s\right) 
\]
where we assume that the manifold $M$ on which we assume the theory to be
defined has been approximated by a triangulation $\Delta $ and $s$ and $f$
are indices running over the 3-volumina and the two dimensional faces of the
dual lattice $\Delta ^{*}$ of $\Delta $. $B_s$ is a Lie algebra valued
variable for every 3-volume $s$ and attaching to every face $f$ an element $%
U_f$ from $SU\left( 2\right) $, we define 
\[
U_s=U_{f_1}...U_{f_n}
\]
for $f_1,...f_n$ the faces around the closed 3-volume $s$. Implicitly, this
definition is dependent on a choice of ordering for the faces around a
3-volume and checking for the independence of this ordering is much more
delicate than in the $BF$ theory case (where this is trivial), see the
results in \cite{BM}. So, the main difference - as compared to the $BF$
theory case - is, so far, the replacement of edges by faces and of faces by
3-volumina. We proceed, now, in analogy to \cite{LPR}. Introducing the
partition function 
\[
Z_M=\int dB_sdU_f\ e^{iS\left( B_sU_f\right) }
\]
one verifies that 
\[
Z_M=\int dU_f\ \Pi _s\delta \left( U_{f_1}...U_{f_n}\right) 
\]
by observing that the original integral involved the Fourier transform of
the $\delta $-function. Since those formulae of \cite{LPR} which involve
only $SU\left( 2\right) $ representation theory remain valid for the $BH$
theory case, of course, we can directly conclude (since this step involves
basically the Peter-Weyl theorem) that 
\[
Z_M=\int dU_f\ \Pi _s\sum_j\left( 2j+1\right) Tr_j\left( U_s\right) 
\]
and by exchanging sum and integral that 
\[
Z_M=\sum_{j_s}\int dU_f\Pi _s\left( 2j_s+1\right) Tr_{j_s}U_s
\]
where $j_s$ is a colouring of 3-volumes by spins $j$ and the sum is taken
over all colourings.

Let us introduce additional variables $v$ and $l$, running over the nodes
and the edges of $\Delta ^{*}$, respectively. In addition let $S,F,L,V$ be
the total number of 3-volumina, faces, edges, and nodes of $\Delta ^{*}$,
respectively. Since formula (21) of \cite{LPR} is a formula which is only
dependent on properties of $SU\left( 2\right) $, again, we can conclude that
adjacent faces - and therefore all faces - have to carry one and the same
representation of $SU\left( 2\right) $. Since each edge in the dual of a
triangular lattice has three faces attached to it, we get combinatorially
the same structure as in the case of $BF$ theory with the replacement 
\begin{eqnarray*}
f &\rightarrow &s \\
l &\rightarrow &f \\
v &\rightarrow &l
\end{eqnarray*}
But while vertices are separated, this is not true for the edges, i.e. on
counting the number of traces which we have to take in the foregoing
formula, we have to count from the different edges attached to the same
closed face only one at a time in order to avoid overcounting. In
consequence, the number of traces we get is given by $L-V$ instead of $L$
(as might be suggested by naively applying the above replacement rule). So,
we get 
\[
Z_M=\sum_j\left( 2j+1\right) ^{S-F+L-V} 
\]
which is a topological invariant of $M$ normally called the Euler
characteristic of $M$ in the literature. This result generalizes the fact
that the partition function of two dimensional $SU\left( 2\right) $ $BF$
theory is determined by the Euler number of the background manifold to the
three dimensional case.

\bigskip

We now treat three dimensional $BH$ theory in complete analogy to the
spinfoam model approach to two dimensional $BF$ theory. For an introduction
to and details of the spinfoam model approach we refer the reader to \cite
{LPR}.

Let $\Phi $ be a real valued scalar field on $SU\left( 2\right) \times
SU\left( 2\right) $ which is symmetric in all its variables and has right $%
SU\left( 2\right) $ invariance, i.e. for all $g,g_1,g_2\in SU\left( 2\right) 
$%
\[
\Phi \left( g_1,g_2\right) =\Phi \left( g_2,g_1\right) 
\]
and 
\[
\Phi \left( g_1,g_2\right) =\Phi \left( g_1g,g_2g\right) 
\]
Define an action $S\left( \Phi \right) $ by 
\begin{eqnarray*}
S\left( \Phi \right) &=&\int_{SU\left( 2\right) \times SU\left( 2\right)
}dg_1dg_2\Phi ^2\left( g_1,g_2\right) \\
&&+\frac \lambda {3!}\int_{\left( SU\left( 2\right) \right)
^3}dg_1...dg_3\Phi \left( g_1,g_2\right) \Phi \left( g_2,g_3\right) \Phi
\left( g_3,g_1\right)
\end{eqnarray*}
It can be shown that the action $S\left( \Phi \right) $ generates through
its Feynman diagrams a spinfoam model for two dimensional $BF$ theory with $%
SU\left( 2\right) $ connection and therefore generates quantized two
dimensional $BF$ theory with $SU\left( 2\right) $ connection in a way which
is not dependent on a fixed background topology and includes a natural
prescription for the sum over topologies in the path integral.

The action $S\left( \Phi \right) $ consists of two parts: The kinetic term
which is quadratic in the field and the interaction term. The structure of
the action is determined geometrically in the following way: The number of
variables of the field is given by the dimension $d$, the number of
integrals of the interaction term is just the number of edges of the $d$%
-simplex and the number of fields in the interaction term is given by the
number of nodes of the $d$-simplex where the edges which meet at a given
node give the arguments of the fields if we assume the integration variables
of the interaction term to be attached to the edges. E.g. for four
dimensional $BF$ theory, $\Phi $ is a real valued right $SU\left( 2\right) $
invariant scalar field on $SU\left( 2\right) ^4$ which is symmetric in all
its variables, the interaction term involves ten integrals and is of fifth
degree in $\Phi $.

Let us now generalize this approach to the case of $SU\left( 2\right) $ $BH$
theory in three dimensions. Let $\Omega $ be a right $SU\left( 2\right) $
invariant real valued scalar field on $SU\left( 2\right) \times SU\left(
2\right) $ which is symmetric. We begin by constructing the interaction term 
$S_I\left( \Omega \right) $. A natural guess is that the number of integrals
in $S_I\left( \Omega \right) $ should be given by the number of faces of the
tetrahedron while the degree in $\Omega $ should be just the number of edges
of the tetrahedron with the arguments labeling the two faces meeting at each
edge. So, with $g_1,...,g_4$ giving a colouring of the faces of the
tetrahedron by $SU\left( 2\right) $ elements, we define 
\begin{eqnarray*}
S_I\left( \Omega \right) &=&\int_{SU\left( 2\right) ^4}dg_1...dg_4\ \Omega
\left( g_1,g_2\right) \Omega \left( g_2,g_4\right) \Omega \left(
g_2,g_3\right) \\
&&\Omega \left( g_1,g_3\right) \Omega \left( g_1,g_4\right) \Omega \left(
g_3,g_4\right)
\end{eqnarray*}
Next, we consider the kinetic term $S_K\left( \Omega \right) $. The fact
that the kinetic term of $BF$ theory is quadratic in the field is linked to
the fact that the 1-form connection $A$ propagates along edges of $\Delta
^{*}$ and therefore each propagator has two endpoints. For a 2-form
connection the number of endpoints of the propagator is given by the number
of edges around a closed face and therefore 
\[
S_K\left( \Omega \right) =\int_{SU\left( 2\right) ^2}dg_1dg_2\left(
\sum_k\Omega ^k\left( g_1,g_2\right) \right) 
\]
where the sum runs over the possible numbers of edges of a closed face in $%
\Delta ^{*}$. The complete action $S\left( \Omega \right) $ for three
dimensional $BH$ theory is then defined as 
\[
S\left( \Omega \right) =S_K\left( \Omega \right) +S_I\left( \Omega \right) 
\]

\bigskip

\begin{remark}
Comparing to the definition of connections and curvature on gerbes in \cite
{BM} one might on first sight assume that besides $\Omega $ one should
introduce a field $\Phi $ on $Aut\left( SU\left( 2\right) \right) $ which
behaves much in the way as the field in the $BF$ theory case. While $\Omega $
corresponds to the transition elements $g_{ijk}$ of a gerbe in \cite{BM},
the field $\Phi $ would correspond to the $\lambda _{ij}$. But observe that
the $\lambda _{ij}$ are not independent of the $g_{ijk}$ and one therefore
has to restrict to a single field $\Omega $ in the action in order to have
the action corresponding to a properly defined configuration space.
\end{remark}

\bigskip

One now checks that the action $S\left( \Omega \right) $ leads with the
partition function 
\[
Z=\int d\Omega \ e^{-S\left( \Omega \right) } 
\]
to the amplitudes 
\[
A_{\Delta ^{*}}=\int dg_e\ \Pi _s\delta \left( g_{f_1}...g_{f_n}\right) 
\]
where $f_1,...,f_n$ are, again, the faces around a closed 3-volume $s$ and
further generates three dimensional $BH$ theory.

\bigskip

\section{Seven dimensions}

In order to generalize our findings to the higher dimensional case, we
collect the following data on $d$-dimensional simplices: We have 
\begin{eqnarray*}
V &=&d+1 \\
L &=&\left( 
\begin{array}{c}
d+1 \\ 
2
\end{array}
\right) =\frac d2\left( d+1\right) \\
F &=&\left( 
\begin{array}{c}
d+1 \\ 
3
\end{array}
\right) =\frac d6\left( d+1\right) \left( d-1\right) \\
N_{F,L} &=&d-1
\end{eqnarray*}
where $N_{F,L}$ denotes the number of faces that meet at an edge.

Let $\Delta $ be a triangulation of a seven dimensional manifold and $\Delta
^{*}$ the dual, again. We define $S_K\left( \Omega \right) $ as above from $%
\Delta ^{*}$ where $\Omega $ is now a right $SU\left( 2\right) $ invariant
real valued scalar field on $\left( SU\left( 2\right) \right) ^6$ which is
symmetric in all its variables. Then, $BH$ theory in seven dimensions has to
be generated in a topology independent way by an auxiliary field theory with
the action 
\[
S\left( \Omega \right) =S_K\left( \Omega \right) +S_I\left( \Omega \right) 
\]
with 
\[
S_I\left( \Omega \right) =\int_{\left( SU\left( 2\right) \right)
^{56}}dg_1...dg_{56}\ \Omega \left( ...\right) ...\Omega \left( ...\right) 
\]
where $S_I\left( \Omega \right) $ is of order 28 in $\Omega $ and the
arguments of the factors of $\Omega $ label the six faces meeting at an edge
of the 7-simplex.

Observe that $BH$ theory in seven dimensions includes as a special case the
action 
\[
S\left( G\right) =\int Tr\left( G^2dG\right) 
\]
for a 2-form connection $G$ which gives the simplest possible generalization
of three dimensional Chern-Simons theory to a seven dimensional theory with
a 3-form field strength. The above action $S\left( \Omega \right) $ for an
auxiliary field theory which generates $BH$ theory in seven dimensions might
therefore be of interest as a simple model to begin to study properties of
the seven dimensional topological field theories, arising in $M$-theory, in
a topology independent way.

\bigskip

\bigskip

\textbf{Acknowledgements:}

K.G.S. thanks the Deutsche Forschungsgemeinschaft (DFG) for support by a
research grant and the Erwin Schr\"{o}dinger Institute for Mathematical
Physics, Vienna, for hospitality.

\end{document}